\documentstyle[12pt]{article}
\input epsf
\begin{document}

\begin{flushright}
RU--96-37 \\
hep-ph/9608330\\
January 24, 1997\\
\end{flushright}

\begin{center}
\bigskip\bigskip\bigskip\bigskip

{\Large \bf Light Gluino Mass and Condensate\\ from Properties of 
$\eta$ and $\eta'$}

\vspace{0.5in}

{\bf Glennys R. Farrar and G.T. Gabadadze}

\vspace{0.2in}

{\baselineskip=14pt
Department of Physics and Astronomy, Rutgers University,\\
Piscataway, New Jersey 08855-0849, USA}
\vspace{0.2in}

\vspace{1.5cm}

{\bf Abstract}
\end{center}
We investigate whether known properties of the $\eta'$ meson are
consistent with its being the Goldstone boson of the spontaneously
broken anomaly-free $R$ symmetry required in the light gluino
scenario.  We fit the masses and $2\gamma$ decays of
the $\eta$ and $\eta'$ mesons, and also their production in radiative
$J/\psi$ decays.  We find that the $\eta-\eta'$ system is
well-described in the light gluino scenario, if 
$m_\lambda\simeq (84-144)~{\rm MeV}$ and 
$\langle{\bar\lambda}\lambda\rangle \simeq -(0.15-0.36) ~{\rm GeV}^3$.  
These values are in the range expected when the gluino gets
its mass entirely from radiative corrections.

\vspace{1.5cm}



\newpage
\vspace{1.5cm}

It has been  known for a long time that there exists a window
for a long-lived light gluino in SUSY  phenomenology
\cite{Farrar84,f:95}. Dimension-3 SUSY breaking operators (including
gaugino masses) may naturally be highly suppressed
in the MSSM \cite{f:99_101}, with a number of attractive
consequences such as eliminating the CP problem of the
MSSM \cite{f:99_101,f:103} and providing the observed dark
matter \cite{f:100}. Gauginos acquire masses through radiative
corrections from electroweak and top-stop loops \cite{bgm,f:96}.  Estimates 
of the constrained soft SUSY breaking parameter space lead to the
gluino mass range $m_\lambda\simeq(0.1 - 1)$ GeV \cite{f:99_101}. Such
a light gluino leads to a  slower running of the strong coupling
constant and might help to resolve the strong coupling constant
discrepancy problem (see \cite{Shifman} and refs.
therein).

The Lagrangian of supersymmetric QCD after decoupling heavy squark
modes consists of the usual QCD plus light gluinos.  Hence, among the
usual symmetries of QCD there exists an additional chiral symmetry
associated with the chiral $U(1)$ transformation of the gluino field.
A particular linear combination of the currents associated with the
quark and gluino chiral rotations is free of the gluon anomaly.  We will call
this anomaly-free symmetry $R$ invariance, and the corresponding
current the $R$ current.  The $R$ invariance is spontaneously
broken by quark and  gluino condensates.  It was argued in ref.
\cite{EVS} that supersymmetry must be explicitly broken by a gluino
mass of order 1 GeV or more, in order to avoid an unobserved Goldstone
boson.   However it was suggested in ref. \cite{Farrar84} that the
required pseudogoldstone boson might be identified with the $\eta'$; a naive
estimate for the required gluino condensate was given in ref.
\cite{f:99_101}. In this scenario the  $0^{-+}$ state which
gets its mass from the anomaly is expected \cite {f:95} to have mass
approximately  $1.5$ GeV and is naturally identified as
the otherwise-mysterious ''extra'' singlet pseudoscalar at $1410$ MeV
\cite {ClFarLi}. Approximately degenerate with it is a spin-1/2 
''glueballino'' whose detection is discussed in ref. \cite {f:prl}.

Here we investigate in greater detail the possibility that the $\eta'$
is the required pseudogoldstone boson of spontaneously broken
R-invariance.  In addition to giving a more accurate determination of the
gluino condensate, we examine whether a non-vanishing gluino condensate
is consistent with the $\gamma \gamma$ decay rates of $\eta$ and $\eta'$, and
with the relative production rates of $\eta$ and $\eta'$ in $J/\psi$
radiative decays.  As in the theory without gluinos, the $\eta,~\eta'$
system must be analyzed together, since mixing between flavor octet
and singlet components is essential for understanding their $\gamma
\gamma$ widths. A basic tool in our consideration is the QCD
sum rule method \cite{SVZ} (reviewed in refs. \cite{RRI},\cite{N},
\cite{S}).   

The paper is organized as follows. We begin by fixing the
conventions for the anomaly free axial currents and
mixing angles. After this, we
derive expressions for the singlet and octet pseudoscalar meson decay
constants using the sum rule technique.  Next we consider the
$2\gamma$ decays of $\eta$ and $\eta'$ mesons and derive sum rules
for $J/\psi$ decays into $\eta(\eta')\gamma$.  Finally, imposing all
available experimental restrictions on these processes, we numerically
solve our system of equations, leading to estimates for the light
gluino mass and condensate.

Let us start with the definition of  the anomaly free axial currents
\begin{equation}
J_{\mu 5}^{(8)}={1\over \sqrt{6}} ({\bar u}\gamma_\mu \gamma_5u+{\bar d}
\gamma_\mu \gamma_5d-2{\bar s}\gamma_\mu \gamma_5s) 
\nonumber
\end{equation}
\begin{equation}
J_{\mu 5}^{(0)}={1\over \sqrt{3}} ({\bar u}\gamma_\mu \gamma_5u+{\bar d}
\gamma_\mu \gamma_5d+{\bar s}\gamma_\mu \gamma_5s-{1\over 2 }
{\bar \lambda}\gamma_\mu \gamma_5 \lambda),
\nonumber
\end{equation}
where the Majorana spinor $\lambda$ denotes the gluino field. 
The derivatives of these currents play the role of interpolating operators
for pure octet and singlet pseudoscalar meson states.  The physical
$\eta$ and $\eta'$ mesons are defined through the mixing angle
$\theta$ for which we choose the standard parametrization  
$
|\eta\rangle=|\eta^8\rangle cos\theta-|\eta^0\rangle sin\theta,~~
|\eta'\rangle=|\eta^0\rangle cos\theta+|\eta^8\rangle sin\theta.
$
Following ref. \cite{AkhFre}, define the decay constants $F_0$ and
$F_8$ as 
\begin{eqnarray}
\langle 0|\partial_\mu J_{\mu 5}^{(0)}|\eta'\rangle=F_0~cos\theta~
m_{\eta'}^2,~~~\langle 0|\partial_\mu J_{\mu 5}^{(0)}|\eta\rangle
=-F_0~sin\theta~m_{\eta}^2, \nonumber \\ 
\langle 0|\partial_\mu J_{\mu 5}^{(8)}|\eta'\rangle=F_8~sin\theta~
m_{\eta'}^2,~~~\langle 0|\partial_\mu J_{\mu 5}^{(8)}|\eta\rangle
=F_8~cos\theta~m_{\eta}^2.
\nonumber 
\end{eqnarray}
Consider now  the correlator of two axial currents 
\begin{equation}
q^{\mu}A_{\mu\nu}=iq^{\mu}\int e^{iqx}\langle 0| T J_{\mu 5}^{(0)}(x)
J_{\nu 5}^{(0)}(0)|0\rangle d^4x=-q_{\nu}\Pi(q^2=-Q^2).
\label{corAA}
\end{equation}
In accordance with the QCD sum rule approach \cite{SVZ} the Borel
transforms of the phenomenological and theoretical parts of the
correlator are equated, making the standard decomposition 
\begin{eqnarray}
Im \Pi^{phen}(s)=Im \Pi^{poles}(s)+Im \Pi^{pert}(s)\theta
(s-s_0),\nonumber \\
Im \Pi^{theor}(s)=Im \Pi^{pert}(s)+Im \Pi^{cond}(s).
\nonumber
\end{eqnarray}
The superscripts ``$poles$'', ``$pert$'', and ``$cond$'' label
the resonance, perturbative and power (condensate) contributions
to the correlator.   The result is
\begin{equation}
\label{borel}
\int_0^{\infty} e^{-s\over M^2}Im\Pi^{poles}(s)ds=
\int_0^{s_0} e^{-s\over M^2}Im\Pi^{pert}(s)ds+
\int_0^{\infty} e^{-s\over M^2}Im\Pi^{cond}(s)ds.
\end{equation}
The conventional parametrization for the pole and the condensate
contributions are the following:
\begin{equation}
Im \Pi^{poles}(s)=\pi \sum_{i=1}^{\infty}c_i \delta (s-m_i^2),~~~~
Im \Pi^{cond}(s)=\pi \sum_{j=0}^{\infty}k_j \delta^{(j)} (s)/{j !},
\nonumber
\end{equation}
with $c_i$'s being pole residues, $k_j$'s being gauge invariant
condensates and $\delta^{(j)}$'s denoting the $j$'th 
derivatives of the Dirac's delta function. Substituting these 
expressions into  eq. (\ref{borel}) and expanding in inverse powers
of the Borel parameter (for $M^2>>s_0$) we get the sum rules in each 
order of $1/M^2$.
For practical calculation it is enough to keep 
only the first two equations of this tower. These equations look like
\begin{equation}
\label{sr1}
\pi (c_1+c_2)=\int_0^{s_0}Im\Pi^{pert}(s)ds+\pi k_0,
\end{equation}
\begin{equation}
\label{sr2}
\pi (c_1m_1^2+c_2m_2^2)=\int_0^{s_0}s Im\Pi^{pert}(s)ds-\pi k_1.
\end{equation}
It is not difficult to see that these relations  are nothing but the QCD
finite energy sum  rule (FESR) \cite{FESR},
modified by the condensate contributions \cite{FESRM}. 

Having set up the general framework, let us turn to the calculations
for our application.  In the case at hand, the one loop calculation
for the singlet correlator in eq. (4) leads to the following
results\footnote{We hereafter neglect the $u$ and $d$ quark masses.}: 
\begin{eqnarray}
Im\Pi^{pert}=\Bigl({m_s^2\over 2\pi}+{m_\lambda^2 \over 3
\pi}\Bigr)+O(\alpha_s m_s^2, \alpha_s m_\lambda^2);
\nonumber \\
k_0=-{2 m_s\langle {\bar s}s\rangle \over 3}-{m_\lambda\langle 
{\bar \lambda}\lambda\rangle \over 6};~~~~~
k_1=-({m_s^2\over 6}+m_\lambda^2)\langle{\alpha_s\over \pi}
G_{\mu\nu}^2\rangle.
\nonumber
\end{eqnarray}
The  constants $c_1,~c_2~, m_1$ and $m_2$ entering the
phenomenological part of the sum rules are: 
$
c_1=F_0^2~ m_{\eta}^2~sin^2\theta,~~~m_1^2=m_{\eta}^2;~~~~
c_2=F_0^2~ m_{\eta'}^2~cos^2\theta,~~~m_2^2=m_{\eta'}^2
$.  Substituting all these relations into  eqs. (\ref{sr1}) and (\ref{sr2}),
determining then $s_0$ from  eq. (\ref{sr2}), and plugging  back its value
into eq. (\ref{sr1}), we arrive at  the following expression for 
the singlet axial constant:
\begin{eqnarray}
\label{F0}
F_0^2=\left( {m_s^2\over 2\pi^2}+{m_\lambda^2\over 3\pi^2}\right)
{(1+0.1~tan^2\theta)(1+ tan^2\theta)\over (1+0.33~
tan^2\theta)^2}\times \nonumber \\
\left(1+\sqrt{1-{2 m_s\langle {\bar s}s\rangle /3+ m_\lambda\langle 
{\bar \lambda}\lambda\rangle /6 \over m_{\eta'}^2 
({m_s^2\over 4\pi^2}+{m_\lambda^2\over 6\pi^2})}{(1+ 0.33~tan^2\theta)\over
(1+0.1~tan^2\theta)}-g }\right)&-& \nonumber \\
\Bigl({2 m_s\langle {\bar s}s\rangle \over 3 m_{\eta'}^2}+{m_\lambda\langle
{\bar \lambda}\lambda\rangle \over 6 m_{\eta'}^2} \Bigr){(1+
tan^2\theta)\over(1+0.33~ tan^2\theta)},
\end{eqnarray}
where
$$
g={({m_s^2\over
6}+m_\lambda^2)\langle{\alpha_s\over \pi}
G_{\mu\nu}^2\rangle \over m_{\eta'}^4 
({m_s^2\over 4\pi^2}+{m_\lambda^2\over 6\pi^2})}{(1+ 0.33~ tan^2\theta)^2\over
(1+0.1~ tan^2\theta)^2}.
$$
Using the results of the above calculation, it is easy to obtain the
analogous sum rule for the decay constant $F_8$. One just needs to
drop the gluino contribution and choose the proper normalization
for the corresponding axial current. The result is:
\begin{eqnarray}
\label{F8}
F_8^2= {m_s^2\over \pi^2}
{(1+9.4~ tan^2\theta)(1+ tan^2\theta)\over (1+3.1~
tan^2\theta)^2} 
\left(1+\sqrt{1-{8 \pi^2 \langle {\bar s}s\rangle  
\over 3 m_{\eta}^2 
m_s} {(1+ 3.1~ tan^2\theta)\over
(1+9.4~ tan^2\theta)}-r}\right) \\ 
-\Bigl({4 m_s\langle {\bar s}s\rangle \over 3 m_{\eta}^2}\Bigr){(1+
tan^2\theta)\over(1+3.1~tan^2\theta)}, \nonumber
\end{eqnarray}
where
$$
r={2\pi^2\over 3}{\langle{\alpha_s\over \pi}
G_{\mu\nu}^2\rangle \over m_{\eta}^4 }{(1+ 3.1~ tan^2\theta)^2\over
(1+9.4~ tan^2\theta)^2}.
$$
Generally speaking there are five unknowns in these equations. 
These are the decay constants $F_0$ and $F_8$,\footnote{Though $F_8$
is known from chiral perturbation theory \cite{DHL}, we include it in
the list of unknowns to provide a consistency check of our
calculations.} mixing angle $\theta$, gluino mass $m_\lambda$ and the
gluino condensate $\langle {\bar \lambda}\lambda\rangle$.  Our
strategy hereafter will be to impose the restrictions on these
unknowns coming from the  $2 \gamma$  decays of the $\eta$ and $\eta'$
mesons and from the radiative decays of the $J/\psi$. 

Let us turn first to the $2\gamma$ decays of the $\eta$ and $\eta'$
mesons.  These decays are governed by the electromagnetic axial
anomaly.  Since gluinos do not interact with photons in leading order,
the gluino part of the singlet current just drops out from the
calculation of the corresponding triangle graphs.  Therefore the
formal expressions for the decay widths are the same as in conventional
QCD.  However the interpolating current for the $\eta'$ meson is now
free of the gluon anomaly and one need not worry about modification 
of PCAC as is
necessary in QCD \cite{ShoreVeneziano}. (Because this current is
already modified by subtracting the proper amount of gluinos).  Taking
into account mixings, one obtains 
\begin{eqnarray}
\label{etagg}
{\Gamma(\eta\rightarrow 2 \gamma)\over\Gamma(\pi\rightarrow 2
\gamma)}=2 \Bigl({m_{\eta}^3\over m_{\pi}^3}\Bigr)\Bigl({F_\pi
cos\theta\over F_8 \sqrt 6}-{2 F_\pi sin\theta\over F_0\sqrt 3}
\Bigr)^2 \\
\label{etaprgg}
{\Gamma(\eta'\rightarrow 2 \gamma)\over\Gamma(\pi\rightarrow 2
\gamma)}=2 \Bigl({m_{\eta'}^3\over m_{\pi}^3}\Bigr)\Bigl({2F_\pi
cos\theta\over F_0 \sqrt 3}+{ F_\pi sin\theta\over F_8\sqrt 6}
\Bigr)^2.
\end{eqnarray}
The available experimental data for these decay rates are \cite{PDG}:
$\Gamma(\eta'\rightarrow 2 \gamma)=(4.26\pm 0.62)$ keV,
$\Gamma(\eta\rightarrow 2 \gamma)=(0.51\pm 0.09)$
keV, $\Gamma(\pi\rightarrow 2 \gamma)=(7.74\pm 0.58)$ eV. 

Eqs. (\ref{etagg}) and (\ref{etaprgg}) are derived using PCAC and soft
meson technique.  The latter is a dubious approximation for the
$\eta'$ meson\footnote{ We are grateful to H. Georgi for bringing this
point to our attention.}.  To obtain some estimate of the effect of
the extrapolation from $m_{\eta'} = 0$ to $m_{\eta'} = 958$ MeV, we
apply the interpolation technique of ref. \cite{BrodskyLepage}.
The result is that the $F_0$'s appearing in eq. (\ref{etaprgg}) are
multiplied by the factor $(1-(m_{\eta'}^2/16\pi^2 F_0^2))$.  Results of
both estimates will be reported below.

Now let us turn to the radiative decays of vector quarkonium. In
particular we will  concentrate on $J/\psi\rightarrow
\eta(\eta')\gamma$. The basic properties of these processes were
worked out using the sum rule technique in refs. \cite{Voloshin},
\cite{NSVZ}.  To understand the mechanism which dominates the decay,
it is relevant to look at the quantum numbers of the particles
involved:  $J/\psi (1^{- -})\rightarrow PS (0^{- +})~ \gamma
(1^{P-})$. In general the spatial parity of the emitted photon may be
negative (for ``electric'' transition) or positive  (for ``magnetic''
transition - see for example \cite{Landafshitz}). From the quantum
numbers of the quarkonium and the pseudoscalar meson, the emitted
photon in the present case must have positive spatial parity and
consequently should be the ``magnetic'' type.  Keeping this fact in
mind, the following three step picture for these decays emerges
naturally:  the heavy ${\bar c}c$ quark system being in a vector state
emits the ``magnetic'' photon and turns into a pseudoscalar state of
two $c$ quarks (a virtual $\eta_c$ meson). Then this virtual $\eta_c$
emits two gluons in a pseudoscalar state which hadronize
into the final pseudoscalar meson ($\eta$ or $\eta'$).  Using
the operator $i{\bar c}\gamma_5 c$ for the interpolating field of the
$\eta_c$ meson, one obtains the following ratio:
\begin{equation}
\label{Jpsi}
{\Gamma(J/\psi\rightarrow \eta' \gamma)\over\Gamma(J/\psi\rightarrow
\eta \gamma)}=\Bigl|{\langle 0|i{\bar c}\gamma_5 c|\eta'\rangle \over
\langle 0|i{\bar c}\gamma_5 c|\eta\rangle}\Bigr|^2
\Bigl({m_{J/\psi}^2-m_{\eta'}^2 \over m_{J/\psi}^2-m_{\eta}^2}\Bigr)^3.
\end{equation}
It is convenient to  introduce the notation  
$\langle 0|i{\bar c}\gamma_5 c|\eta'\rangle \equiv
\alpha_s^2
 a_{\eta'}/\pi^2$ and
$\langle 0|i{\bar c}\gamma_5 c|\eta\rangle \equiv \alpha_s^2 a_{\eta}/\pi^2$.
Eq. (\ref{Jpsi}) and the experimental data on these decays \cite{PDG} give
\begin{equation}
\label{aeta}
|a_{\eta'}|\simeq (2.48\pm 0.16) |a_{\eta}|.  
\end{equation}

The next step is to relate $a_\eta$ and $a_\eta'$ to properties of the
$\eta$ and $\eta'$ using sum rules.  Therefore consider the two-point
correlator 
\begin{equation}
P(q^2)=i\int e^{iqx}\langle 0|T i{\bar c}\gamma_5 c(x) \partial_\mu
J_{\mu 5}^0(0)|0\rangle d^4x.
\end{equation}
The calculational procedure is the same one used above for the 
determination of $F_0$ and $F_8$. The phenomenological part is
saturated by the physical $\eta$ and $\eta'$ meson states. The
leading perturbative part is determined by the three-loop diagram
shown in fig.1(a). This diagram has been calculated recently in ref.
\cite{ChetKw}, using an  expansion in inverse powers of  heavy quark
mass \cite{HQ}.  In our case this is the leading approximation in
$1/m_c$. The leading contribution to the nonperturbative part arises
from the diagrams of fig. 1(b,c,d).  We have used the heavy quark mass
expansion and dimensional regularization, with the 't Hooft-Veltman
prescription \cite{tHooftVeltman} for the $\gamma_5$ matrix in
$D=4-2\epsilon$ dimensional space-time.  The calculational technique is
based on the integration by parts method of ref. \cite{ChetKatTkach}.

\begin{figure}
\begin{center}
\vskip -220pt
\epsfxsize=6in
\epsfysize=8in
\epsffile{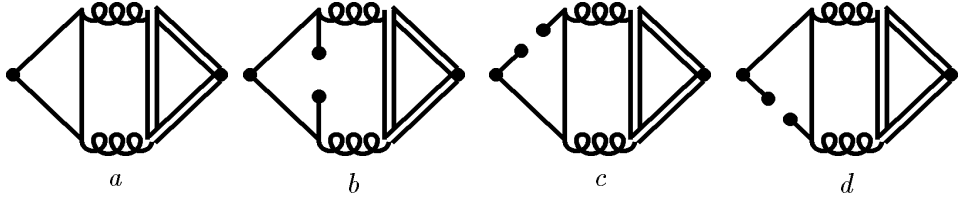}
\vskip -220pt
\caption{ The double plain lines denote the $c$ quark propagators 
and the single plain lines
denote the light quark or gluino propagators.}
\end{center}
\end{figure}
The results corresponding to the three diagrams of figs. 1b, 1c and 1d
are  
$$
Im~P_b(s+i\varepsilon)=0;~~~~~
Im~P_c(s+i\varepsilon)=Im~P_d(s+i\varepsilon)=
$$
$$
{1\over 2\sqrt{3}}
\Bigl({\alpha_s \over \pi}\Bigr)^2
({m_s
\langle{\bar s}s\rangle \over 3 m_c}-{9\over 24}
{m_\lambda\langle{\bar \lambda}\lambda\rangle \over m_c}\Bigr)
\epsilon (2G(1,1)-G(2,1)),
$$
where $G(1,1)=1/\epsilon +ln{4\pi}-\gamma+2+o(\epsilon)$ and 
$G(2,1)=-1/\epsilon-ln{4\pi}+\gamma+o(\epsilon)$ are the $G$-functions
introduced in ref. \cite{ChetKatTkach}, and $\gamma\simeq 0.5772$ is
the Euler constant.  Using these results for the theoretical part, the
sum rules are
\begin{eqnarray}
\label{srF0}
F_0(a_{\eta'}m_{\eta'}^2cos\theta-a_{\eta}m_{\eta}^2sin\theta)=
{2(m_s^2-3 m_\lambda^2)\over m_c \sqrt{3}}
{3\over 8 \pi^2}\int_0^{s_0}f(s)ds &-& \nonumber \\
{1\over \sqrt 3}\bigr(-{m_s
\langle{\bar s}s\rangle \over m_c}+{9\over 8}
{m_\lambda\langle{\bar \lambda}\lambda\rangle \over m_c}\Bigr) s_0,
\end{eqnarray}
\begin{eqnarray}
\label{srF0'}
F_0(a_{\eta'}m_{\eta'}^4cos\theta-a_{\eta}m_{\eta}^4sin\theta)=
{2(m_s^2-3 m_\lambda^2)\over m_c \sqrt{3}}
{3\over 8 \pi^2}\int_0^{s_0}s f(s)ds &-& \nonumber \\
{1\over \sqrt 3}\bigr(-{m_s
\langle{\bar s}s\rangle \over m_c}+{9\over 8}
{m_\lambda\langle{\bar \lambda}\lambda\rangle \over m_c}\Bigr)
{s_0^2\over 2},
\end{eqnarray}
where the function $f(s)=4s-sln{s\over m_c^2}+{s^2\over m_c^2}
({61\over 324}-{7\over 108}ln{s\over m_c^2})$ appears as the 
imaginary part of diagram (1a)  \cite{ChetKw}.  

In analogy with the singlet case one can write down the same sum rules
for the correlator with the nonsinglet current.  Taking the proper
normalization for this last we get
\begin{eqnarray}
\label{srF8}
F_8(a_{\eta}m_{\eta}^2cos\theta+a_{\eta'}m_{\eta'}^2sin\theta)=
{-4m_s^2\over m_c \sqrt{6}}
{3\over 8 \pi^2}\int_0^{s_0^{(8)}}f(s)ds &-& \nonumber \\
{1\over \sqrt 6}{2m_s
\langle{\bar s}s\rangle \over m_c} s_0^{(8)},
\end{eqnarray}
\begin{eqnarray}
\label{srF8'}
F_8(a_{\eta}m_{\eta}^4cos\theta+a_{\eta'}m_{\eta'}^4sin\theta)=
{-4m_s^2\over m_c \sqrt{6}}
{3\over 8 \pi^2}\int_0^{s_0^{(8)}}sf(s)ds &-& \nonumber \\
{1\over \sqrt 6}{2m_s
\langle{\bar s}s\rangle \over m_c}{s_0^{(8)2}\over 2}.
\end{eqnarray}

We now have enough equations to determine the unknown constants. In
fact, we have nine equations (\ref{F0}-\ref{etaprgg}), (\ref{aeta}),
(\ref{srF0}-\ref{srF8'}) and nine unknowns (
$\theta,~F_0,~F_8,m_\lambda,~ \langle{\bar \lambda}\lambda\rangle,
~a_{\eta},$ $a_{\eta'},~ s_0$ and $s_0^{(8)}$) which depend on $m_s$,
$\langle{\bar s}s\rangle$, $\langle{\alpha_s\over
\pi}G_{\mu\nu}^2\rangle$.  For the $s$ quark mass, recent lattice
calculation gives $m_s( 2~GeV)=(141\pm 17)$ MeV \cite{mslat};  QCD sum
rules give $m_s(1~GeV)=(189\pm 32)$ MeV \cite{msj} and
$m_s(1~GeV)=(171\pm 15)$ MeV \cite{msc}.  In our calculations we take
the average of the above three results $m_s(1~GeV)= (167\pm 13)$ MeV.  We
take the $s$ quark condensate from ref. \cite{OP} (see also the recent 
paper \cite{ss}) $\langle{\bar s}s\rangle=(0.7\pm 0.2)\langle{\bar 
u}u\rangle= -(0.011\pm0.003)~{\rm GeV}^3$. There are a number of estimates
for the gluon condensate in the literature (see refs.
\cite{gg},\cite{SVZ}). We take the world average value of these
calculations $\langle{\alpha_s\over \pi} G_{\mu\nu}^2\rangle= (2.5\pm
0.9)~10^{-2}~{\rm GeV}^4$.  

Our procedure is this.  First we use eq. (\ref{etagg}) to find $F_0$
(or $F_0(1-(m_{\eta'}^2/16\pi^2 F_0^2))$ using the interpolation based
on the Brodsky-Lepage analysis) and substitute it into eq. 
(\ref{etaprgg}). After this we have two equations, (\ref{F8}) and
(\ref{etaprgg}), for $F_8$ and $\theta$. Solving this system
numerically we find  $F_8\simeq (130\pm 20(stat.))$ MeV and $\theta
\simeq  -(23.5^0\pm 2.7^0(stat.))$.  The value obtained here for $F_8$
agrees within error bars with the chiral perturbation theory estimate
$F_8\simeq1.25 F_\pi\simeq 116~ $ MeV \cite{DHL}. Our value for
$\theta$ is also in agreement with previous calculations (see for
example ref. \cite{GilKauf}).  This gives confidence in these values
for $m_s$ and  $\langle{\bar s}s\rangle$, and in the present method
combining PCAC and sum rule techniques for calculating the decay
constants.  We find $F_0\simeq (93\pm 17(stat.))$ MeV with the
conventional soft meson extrapolation for the $\eta'$, and $F_0\simeq
(136\pm 13(stat.))$ MeV with the alternative procedure described
above. 

Next we make use of the data and SR's for $J/\psi \to \eta \gamma$ and
$J/\psi \to \eta' \gamma$ , namely eqs.
(\ref{aeta},\ref{srF8},\ref{srF8'}) to determine $a_{\eta},~a_{\eta'}$
and $s_0^{(8)}$ numerically using Mathematica.  The procedure is
trivial but cumbersome and yields:  
$|a_{\eta'}|\simeq (202.1\pm 45.3)10^{-3} {\rm GeV}^2,~~|a_{\eta}|\simeq
(81.6\pm 18.1)10^{-3} {\rm GeV}^2$ and $s_0^{(8)}\simeq (1.7\pm
0.2){\rm GeV}^2.  $
Note that we took from experiment the {\it ratio}
$|a_{\eta}|/|a_{\eta'}|$, but have not obtained their individual
values.  Thus if we had a reliable independent way of estimating the 
matrix element $\langle 0|i{\bar c} \gamma_5 c|\eta_c\rangle \equiv F_c$
appearing in the amplitude for $J/\psi \rightarrow \eta
\gamma$ and $J/\psi \rightarrow \eta' \gamma$, we would have a
prediction for the overall magnitude of these rates.  Unfortunately,
$F_c$ cannot be determined independently with sufficient accuracy to
give us a useful test\footnote{M. Shifman, private communication.}.

The results above allow us to turn to the calculation of the basic
quantities, the gluino mass and condensate.  Substituting the
quantities obtained earlier into eqs. (\ref{F0}, \ref{srF0},
\ref{srF0'}), we have a system of three equations with the last three
unknowns: $m_\lambda,~ \langle{\bar \lambda}\lambda\rangle$ and $
s_0$.  The numerical solution of this set provides us with the
following results using $F_0$ extracted with the conventional soft
meson aproximation: 
$ 
s_0\simeq (1.8\pm 0.2)~{\rm GeV}^2,~~~m_\lambda\simeq (120\pm24)~{\rm
MeV},$ and $ \langle{\bar \lambda}\lambda\rangle \simeq -(0.22 \pm
0.07) ~{\rm GeV}^3$.  Using $F_0$ obtained from the alternative
extrapolation in $m_{\eta'}$ gives:
$ s_0\simeq (1.9\pm 0.2)~{\rm GeV}^2,~~~m_\lambda\simeq (102\pm 
18)~{\rm MeV}$, and $
\langle{\bar \lambda}\lambda\rangle \simeq -(0.31 \pm 0.05) ~{\rm GeV}^3
$.  From this range of results we arrive at the final estimate 
$$m_\lambda\simeq (84 - 144) {\rm MeV},~~~
\langle{\bar \lambda}\lambda\rangle \simeq -(0.15 - 0.36) ~{\rm GeV}^3.
$$

Before concluding, we note that in  conventional QCD the  $\eta'$ is 
identified with the particle which acquires mass due to the axial
anomaly. The Witten-Veneziano (WV) formula \cite{Witten} 
relates the vacuum topological 
susceptibility of pure Yang-Mills theory to the decay constant and mass 
of the $\eta'$ meson in the large $N_c$ limit. (For a recent discussion see
ref.  \cite {SchafShur}).
In the limit when at least one quark is massless this relation looks like
$$
\left ({g^2\over 16\pi^2}\right )^2\int \langle 0|TG^a\tilde G^a(x)
G^a\tilde G^a(0)|0\rangle d^4x|^{YM}_{
N_c\rightarrow \infty}={1\over 3} F_0^2 m_{\eta'}^2|_{N_c\rightarrow
\infty}. \eqno (21)
$$
In the scenario we have considered here, the $\eta'$ is a
pseudogoldstone boson and a different state acquires its mass from the
anomaly.  (In supersymmetric YM theory this state is
a pure gluino-gluino bound state \cite {VenezYank}.)
It is possible  to derive an analogous  formula to eq. (21) in  
the theory with a light gluino. The only difference  is
that the left hand side should be calculated in 
supersymmetric YM theory. The right hand side is unchanged, and still
depends on the
decay constant and mass of the  physical $\eta'$ meson which now
contains a light gluino. 
Unfortunately, the accuracy of the determination of the 
vacuum topological susceptibility (for a review see ref. \cite{SchafShur}) 
is not sufficient  at present to decide   
which of the two scenarios is realized. A detailed discussion  of these and  
related  topics will be presented elsewhere.  

To summarize, in this paper we have undertaken an attempt to identify
the $\eta'$ meson with the Goldstone boson of the spontaneously
broken, anomaly-free $R$ symmetry arising in a theory with a light
gluino.  We have calculated the corresponding decay constant $F_0$
using the QCD sum rule method.  As a consistency check, the analogous
calculation for the octet decay constant $F_8$ is performed. The
result is in good agreement with what was already known before. The
experimental values for the decay rates of $\eta'(\eta)\rightarrow
2\gamma$ and $J/\psi\rightarrow \eta'(\eta)\gamma$ impose certain
restrictions on the values of gluino mass and condensate. Using the
whole system of equations coming from sum rules and from quantities
determined by experiment, we determined the gluino mass and
condensate.  The value of the gluino condensate indicates that the
scale of the breakdown of $R$-symmetry is about 2.5 times larger than 
that of chiral symmetry.  In view of the factor 9/4 larger value of
the adjoint Casimir compared to the fundamental one, this seems to be
of the right magnitude.  Note also that the gluino mass value we have 
found, $m_{\lambda} \sim 80 - 160$ MeV, is in the range predicted in
ref. \cite{f:99_101}.  If the properties of the $\eta,~\eta'$ mesons
had been incompatible with the light gluino scenario, our procedure
would have given nonsensical rather than reasonable values of gluino
mass and condensate.  Thus our result shows a surprising consistency
between the properties of the ground-state pseudoscalar mesons and the 
light gluino scenario, and favors a gluino mass of order 100 MeV. 

The authors are grateful to K.G. Chetyrkin for discussion of the results 
of ref. \cite{ChetKw} and M. Losada for help in using the ``feynMF''
package \cite{feyn}.


\end{document}